\documentclass[11pt]{artikel3}

\usepackage{graphicx,pslatex}
\usepackage{hyperref}

{\catcode`\^^I=13 \globaldefs=1

}

\def\n#{\bgroup \catcode`\$=12 \catcode`\_=12 \catcode`\>=12 \catcode`\<=12 \catcode`\#=12
  \catcode`\&=12 \catcode`\^=12 \catcode`\~=12 \def\\{\char`\\}\relax
  \tt \let\next=}

\newcommand\defined{
  \mathrel{\lower 5pt \hbox{${\equiv\atop\mathrm{\scriptstyle D}}$}}}

\expandafter\ifx\csname definition\endcsname\relax
    
\fi
\expandafter\ifx\csname theorem\endcsname\relax
    
\fi
\expandafter\ifx\csname lemma\endcsname\relax
    
\fi

\newtoks\chaptersourcelist
\newcommand\addchaptersource[1]{
  \edef\temp{\global\chaptersourcelist={\the\chaptersourcelist #1}}\temp
}
\newcommand\listchaptersources{
  \expandafter\ChapterSourceHeader\the\chaptersourcelist\LSR
  \expandafter\ListSourcesRecursively\the\chaptersourcelist\LSR
}
\def\LSR{\LSR}
\def\ChapterSourceHeader#1\LSR{
  \def\test{#1\LSR}
  \ifx\test\LSR
  \else
    \Level 0 {Sources used in this chapter}
  \fi
}
\def\ListSourcesRecursively#1{
  \def\test{#1}
  \ifx\test\LSR
  \else
    \textbf{Listing of code #1}:
    {\footnotesize \verbatiminput{#1}}
    \par
    \expandafter\ListSourcesRecursively
  \fi
}

\usepackage{acronym}
\newwrite\acrowrite
\openout\acrowrite=acronyms.tex
\def\acroitem#1#2{\acrodef{#1}{#2}
    \write\acrowrite{\message{defining #1}\noexpand\acitem{#1}{#2}}
}
\acroitem{AVX}{Advanced Vector Extensions}
\acroitem{BSP}{Bulk Synchronous Parallel}
\acroitem{CAF}{Co-array Fortran}
\acroitem{CUDA}{Compute-Unified Device Architecture}
\acroitem{DAG}{Directed Acyclic Graph}
\acroitem{DSP}{Digital Signal Processing}
\acroitem{FEM}{Finite Element Method}
\acroitem{FPU}{Floating Point Unit}
\acroitem{FFT}{Fast Fourier Transform}
\acroitem{FSA}{Finite State Automaton}
\acroitem{GPU}{Graphics Processing Unit}
\acroitem{HPC}{High-Performance Computing}
\acroitem{HPF}{High Performance Fortran}
\acroitem{ICV}{Internal Control Variable}
\acroitem{MIC}{Many Integrated Cores}
\acroitem{MIMD}{Multiple Instruction Multiple Data}
\acroitem{MPI}{Message Passing Interface}
\acroitem{MTA}{Multi-Threaded Architecture}
\acroitem{NIC}{Network Interface Card}
\acroitem{NUMA}{Non-Uniform Memory Access}
\acroitem{OS}{Operating System}
\acroitem{PGAS}{Partitioned Global Address Space}
\acroitem{PDE}{Partial Diffential Equation}
\acroitem{PRAM}{Parallel Random Access Machine}
\acroitem{RDMA}{Remote Direct Memory Access}
\acroitem{RMA}{Remote Memory Access}
\acroitem{SAN}{Storage Area Network}
\acroitem{SaaS}{Software as-a Service}
\acroitem{SFC}{Space-Filling Curve}
\acroitem{SIMD}{Single Instruction Multiple Data}
\acroitem{SIMT}{Single Instruction Multiple Thread}
\acroitem{SM}{Streaming Multiprocessor}
\acroitem{SMP}{Symmetric Multi Processing}
\acroitem{SOR}{Successive Over-Relaxation}
\acroitem{SP}{Streaming Processor}
\acroitem{SPMD}{Single Program Multiple Data}
\acroitem{SPD}{symmetric positive definite}
\acroitem{SSE}{SIMD Streaming Extensions}
\acroitem{TLB}{Translation Look-aside Buffer}
\acroitem{UMA}{Uniform Memory Access}
\acroitem{UPC}{Unified Parallel C}
\acroitem{WAN}{Wide Area Network}
\acresetall
\closeout\acrowrite

\def\chaptertitle{\csname\chaptername title\endcsname}
\def\chaptershorttitle{\csname\chaptername shorttitle\endcsname}

\usepackage{geometry}
\begin{document}
\title{Performance of MPI sends of non-contiguous data}
\author{Victor Eijkhout}
\date{2018}
\maketitle

\begin{abstract}
  We present an experimental investigation of the performance of MPI
  derived datatypes. For messages up to the megabyte range most
  schemes perform comparably to each other and to manual copying into
  a regular send buffer. However, for large messages the internal
  buffering of MPI causes differences in efficiency. The optimal
  scheme is a combination of packing and derived types.
\end{abstract}

\section{Introduction}

Most \ac{MPI} benchmarks use the send of a contiguous buffer to measure
how close the MPI software gets to the theoretical speed of the
hardware.
However, in practice we often need to send non-contiguous data, such
as the real parts of a complex array, every other element of a grid
during multigrid coarsening, or irregularly spaced elements in a
\ac{FEM} boundary transfer.

To support this, MPI has so-called `derived datatypes' routines, that
describe non-contiguous data. While they offer an elegant interface,
programmers always worry about any performance implications of using
these derived types. In this report we perform an experimental
investigation of  various schemes for sending non-contiguous
data, comparing their performances both to each other, as well as to
the performance of a contiguous send, which we will consider as the
optimal, reference, rate.

Specifically, we investigate the following non-orthogonal issues:
\begin{enumerate}
\item Manually copying irregular data to a contiguous buffer and
  sending the latter, versus using MPI derived datatypes to send irregular data;
\item using both two-sided and one-sided point-to-point; and
\item using buffered and packed sends to counteract adverse effects of
  MPI's internal buffering.
\end{enumerate}

We discuss our schemes in section~\ref{sec:schemes} and our
experimental setup in section~\ref{sec:setup}. Results are graphed and
discussed in section~\ref{sec:results}. We give our overall conclusion
in section~\ref{sec:conclusion}.

\section{Send schemes}
\label{sec:schemes}

We describe the various send schemes that we tested. Any performance
discussion will be postponed to section~\ref{sec:results}.

\subsection{Contiguous send}

To establish a baseline we send a contiguous buffer, and accept the
result as the attainable performance of the hardware/software
combination.

The cost of this scheme is
\begin{enumerate}
\item loading $N$ elements from memory to the processor, and
\item sending $N$ elements through the network.
\end{enumerate}
The two are probably overlapped, and to first order of approximation
we equate memory bandwidth and network transmission speed, so we
assign a proportionality constant of~1 to this scheme.

\subsection{Manual copying}

Since we are sending non-contiguous data, the minimal solution is to
copy the data between a user array and the send buffer. We allocate
the send buffer outside the timing loop, and reuse it.

The cost for this is
\begin{enumerate}
\item the cost of the copying loop, and
\item the cost of the contiguous send.
\end{enumerate}
The copying loop loads $2N$ elements from memory and writes~$N$. The
latter writes can be interleaved with the loads, so most likely only
the loads contribute to the measured time.
Next, the building of the send buffer has to be fully finished before
the \n{MPI_Send} can be issued, so with the assumptions made before,
we expect a slowdown of a factor of~3 over the reference speed.

A further point to note is that the copying loop is of necessity in
user space, so any offload of sends to the \ac{NIC} is of limited
value, reducing the proportionality constant to~2.

\subsection{Derived datatypes}

MPI has routines such as \n{MPI_Type_create_vector} and
\n{MPI_Type_create_subarray} to offer a convenient interface to
non-contiguous data. In a naive implementation, these routines copy
data to an internal MPI buffer, which is subsequently send. In this
case we expect a performance similar to the manual copying scheme; any
degradation indicates a true performance penalty.

However, with enough support of the \ac{NIC} and its firmware, it
would be possible for this scheme to pipeline the reads and sends
similarly to the reference case.
That this is possible in principle was shown in~\cite{LI:MpiDataUMR}.
In practice we don't see this performance,
and our observed performance numbers are
completely in accordance with an assumption of non-pipelining.

\subsection{Buffered sends}

At larger message sizes, it is reasonable to assume that MPI can run
out of internal buffer space, and send the message in packets, giving
a performance degradation. We have investigated whether any benefits
are accrued from using buffered sends. Here a user buffer is attached with
\n{MPI_Buffer_attach} and the send is replaced by \n{MPI_Bsend}.

\subsection{One-sided transfer}

We explored using one-sided point-to-point calls. There are two options:
\begin{enumerate}
\item Transfering the data elements in a loop. However, this requires
  a function call per elements so we did not pursue this.
\item Transfering a single derived type.
\end{enumerate}

While one-sided messages have the potential to be more efficient,
since they dispense with a rendez-vous protocol, they still require
some sort of synchronization. We use \n{MPI_Win_fence}.

\subsection{Packing}

Finally, we considered the use of the \n{MPI_PACKED} data type, used
two different ways.
\begin{itemize}
\item We use a separate \n{MPI_Pack} call for each element. Since this
  uses a function call for each element sent, we expect a low
  performance.
\item We use a single \n{MPI_Pack} call on a derived (vector) datatype.
\end{itemize}

The \n{MPI_Pack} command packs data in an explicitly allocated
send buffer in userspace. This means that we no longer rely on MPI for
buffer management. If the implementation of the pack command is
sufficiently efficient, we expect this scheme to track the manual
copying scheme.

\section{Experimental setup}
\label{sec:setup}

\subsection{Hardware}

We use the following clusters at TACC:
\begin{itemize}
\item Lonestar5: a Cray XC40 with Intel compilers and Cray mpich7.3.
\item Stampede2-knl: a Dell cluster with Knightslanding nodes
  connected with Omnipath, using Intel compilers and Intel MPI.
\item Stampede2-skx: a Dell cluster with dual Skylake nodes connected
  with Omnipath, using Intel compilers and both Intel MPI and mvapich.
\end{itemize}

\subsection{Performance measurement}

We measured the time for 20 ping-pongs, where:
\begin{itemize}
\item The `ping' was the
  non-contiguous send. This used an ordinary \n{MPI_Send} for all
  two-sided versions except \n{MPI_Bsend} for the buffered protocol,
  and \n{MPI_Put} for the one-sided version.
\item In all two-sided schemes the target process executed an \n{MPI_Recv} on a contiguous
  buffer.
\item In the two-sided schemes, there was a zero byte `pong' return
  message. In the one-sided scheme we surrounded the transfer with
  active target synchronization fences; the timers surrounded these fences.
\end{itemize}

Time reported in the figures below is the
total time divided by the number of ping-pongs.
Every ping-pong was timed individually. We used \n{MPI_Wtime}, which
on our platforms had a resolution of $1\cdot10^{-6}$ seconds, and the
minimum measurement ever (for the very smallest message) was around
$6\cdot10^{-6}$ seconds.
Our code is set up to dismiss measurements that are more
than one standard deviation from the average, but in practice this
test is never needed.

All buffers were allocated outside the timing loop,
using 64 byte alignment. Page instantiation was kept outside the
timing loop by setting all arrays explicitly to zero.

In between every two ping-pongs an array of size 50M is
rewritten. This is enough to flush the caches on our systems.

\section{Results and discussion}
\label{sec:results}

\begin{figure}[ht]
  \includegraphics{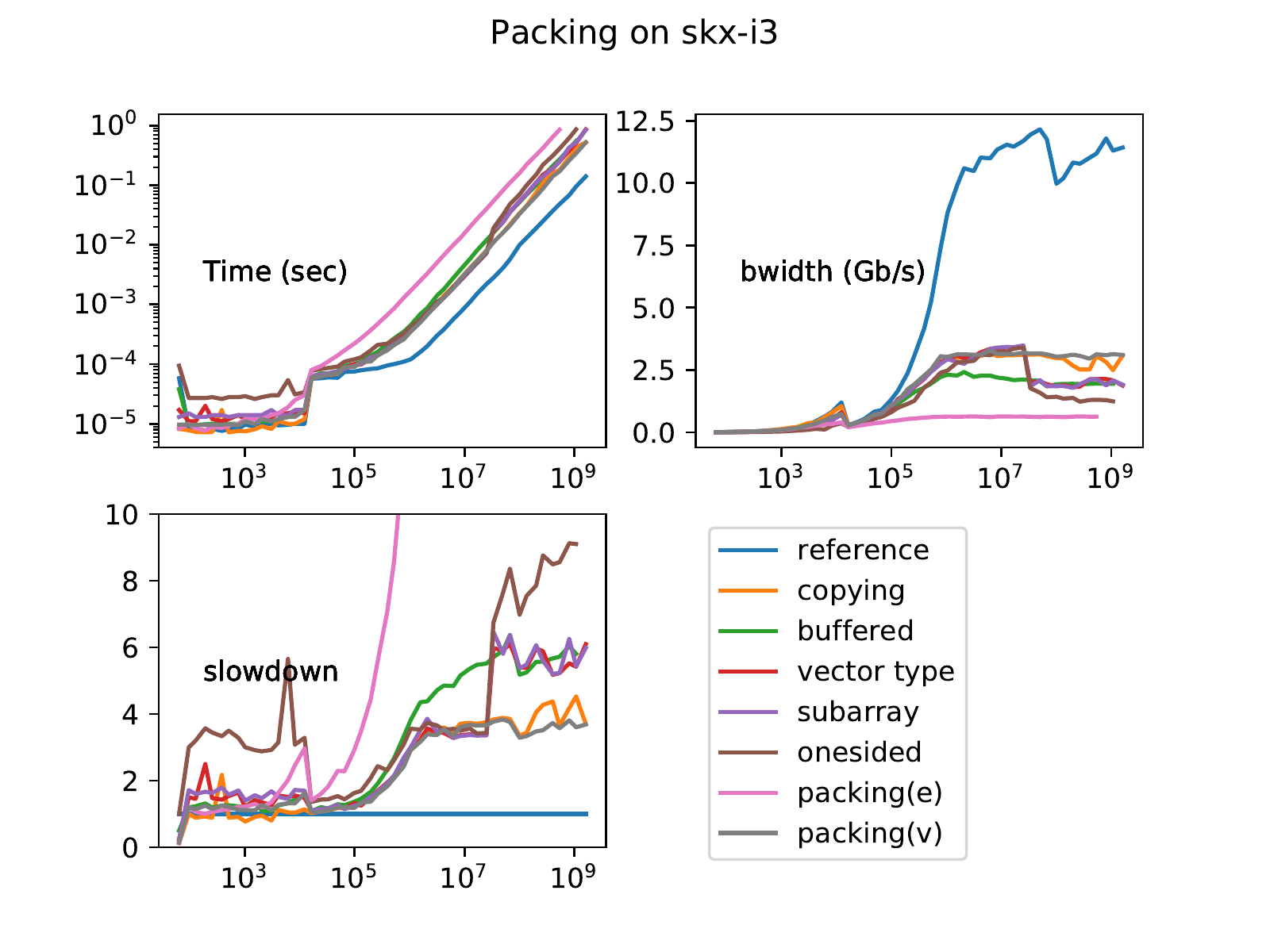}
  \caption{Time and bandwdith on Stampede2-skx using Intel MPI}
  \label{fig:skx-i}  
\end{figure}
\begin{figure}[ht]
  \includegraphics{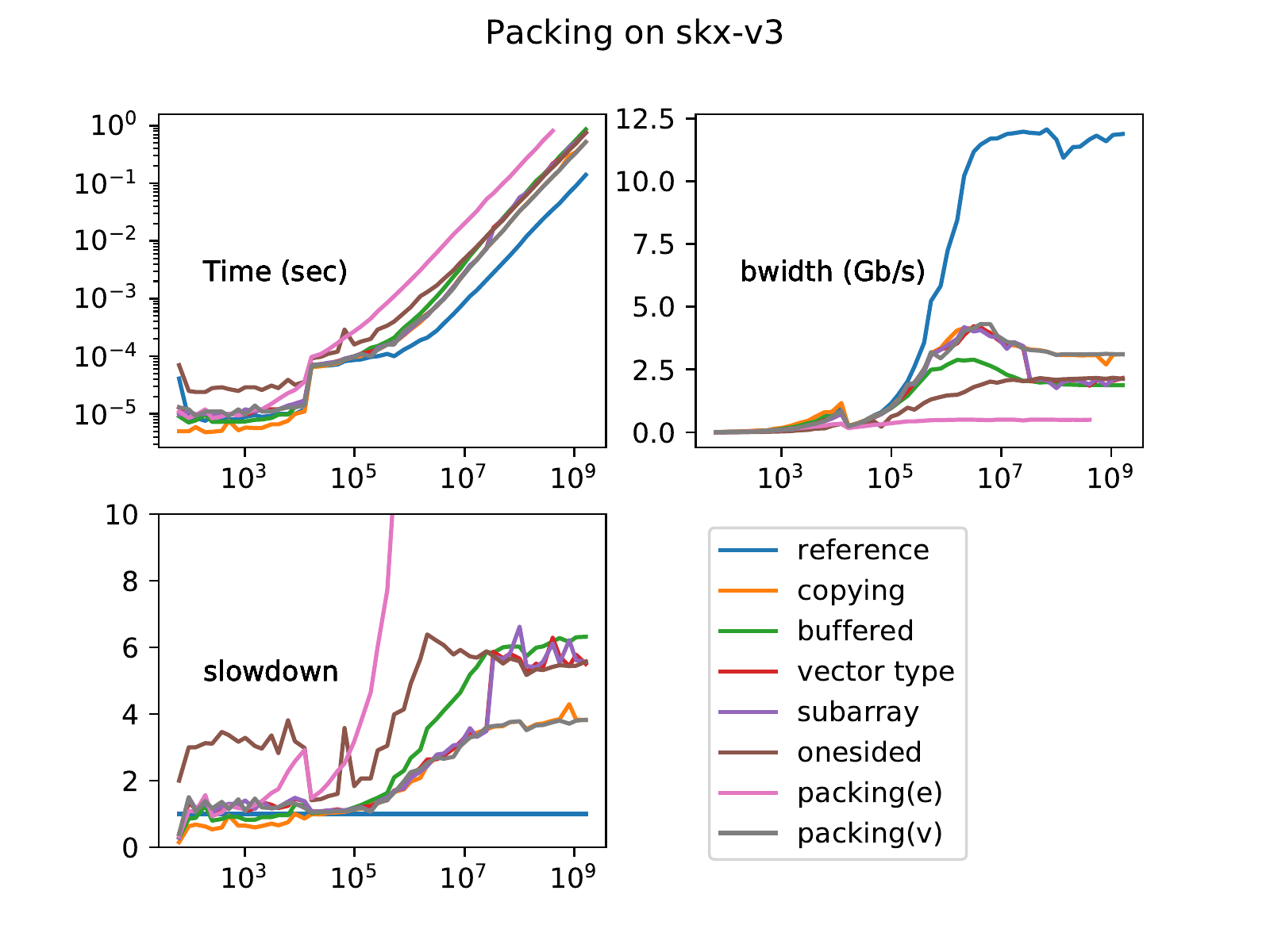}
  \caption{Time and bandwdith on Stampede2-skx nodes using mpivach2}
  \label{fig:skx-v}
\end{figure}
\begin{figure}[ht]
  \includegraphics{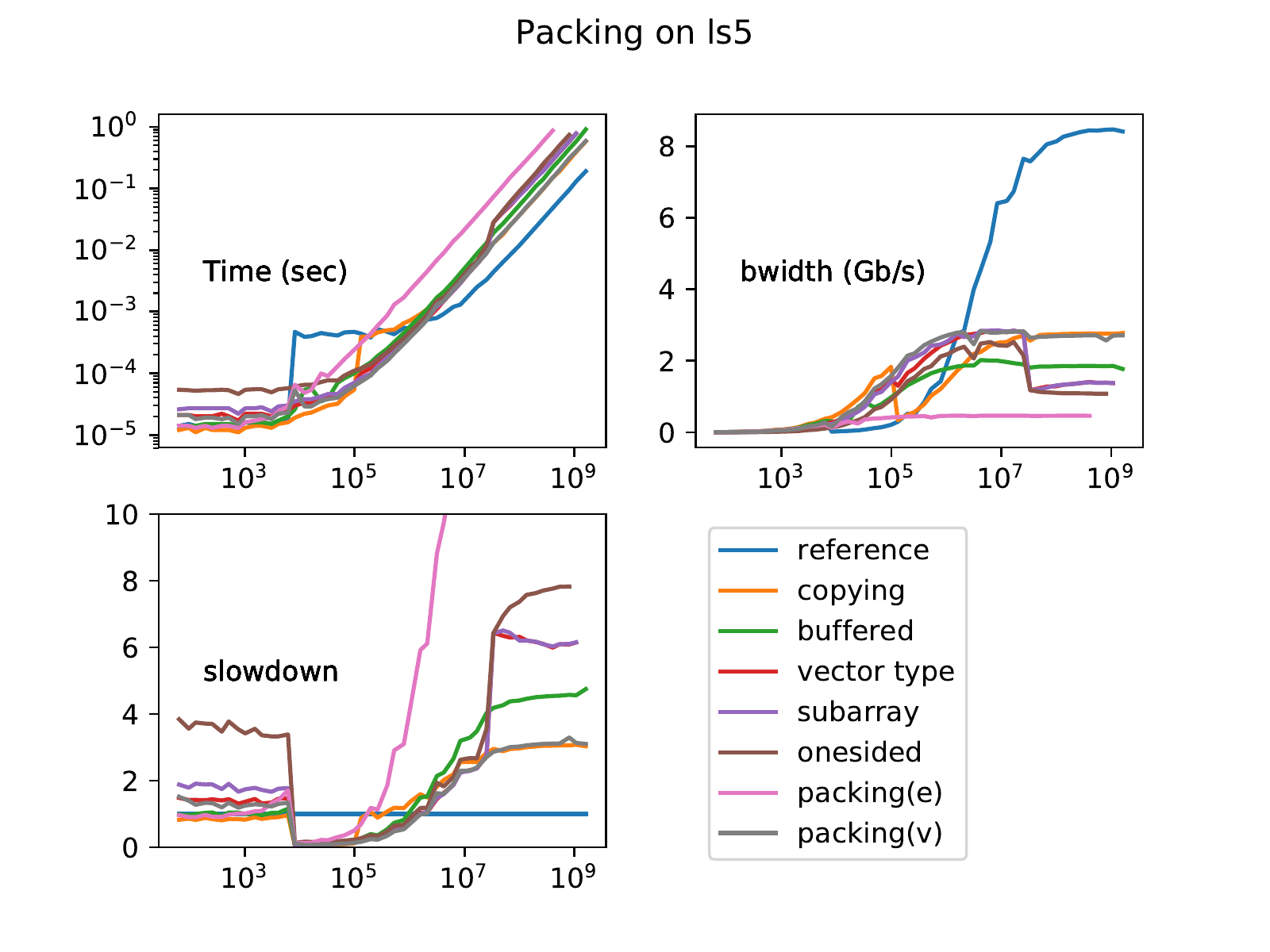}
  \caption{Time and bandwdith on a Cray XC40 using the native MPI}
  \label{fig:ls5}
\end{figure}
\begin{figure}[ht]
  \includegraphics{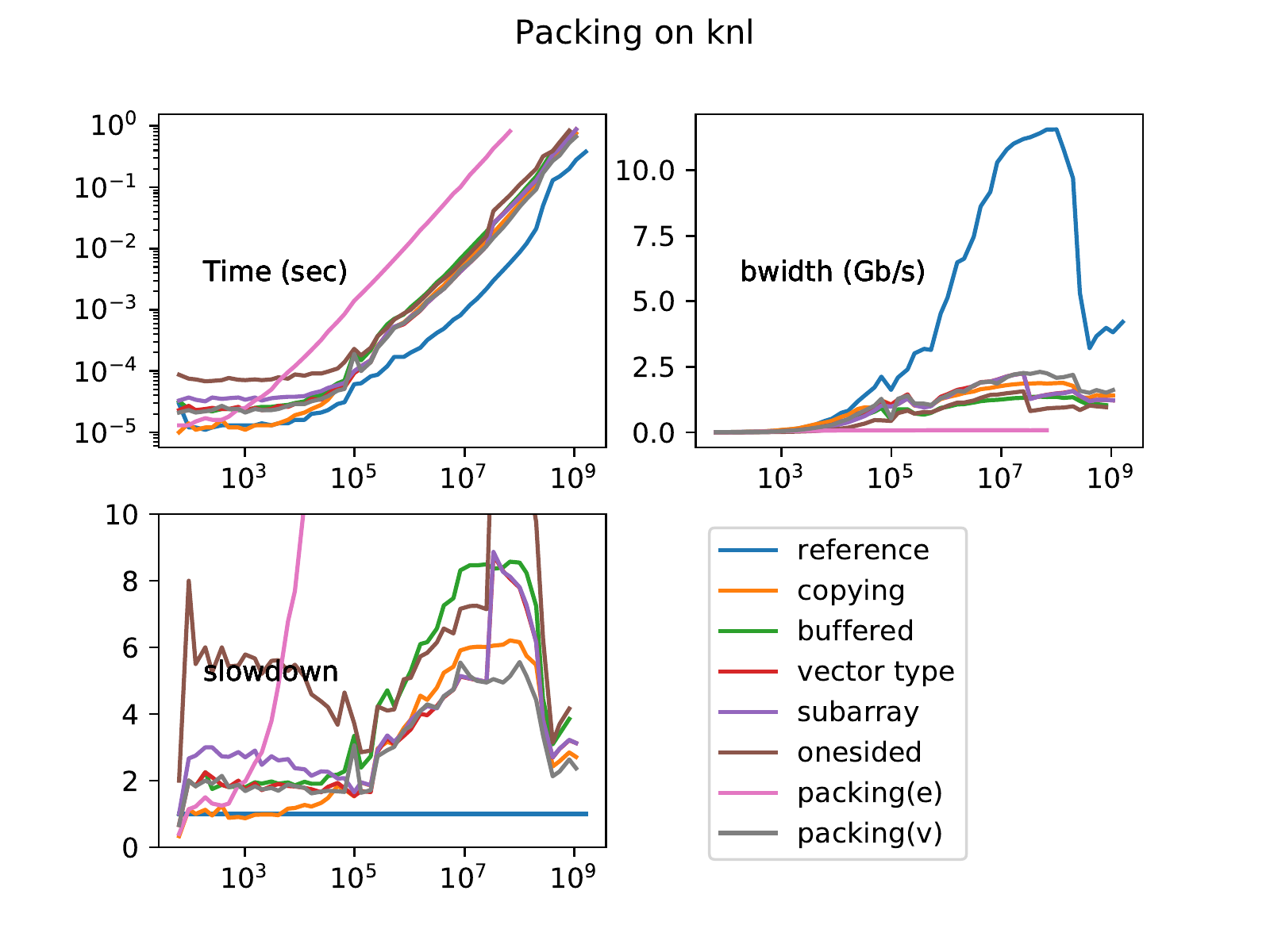}
  \caption{Time and bandwdith on Stampede2-knl using Intel MPI}
  \label{fig:knl}
\end{figure}

We report on the performance with
Skylake processors with Intel MPI (figure~\ref{fig:skx-i}), Skylake
with MVAPICH2 (figure~\ref{fig:skx-v}), Cray with Cray MPICH (figure~\ref{fig:ls5}),
and Knights Landing processors with Intel MPI
(figure~\ref{fig:knl}). The legend on the figures is self-explanatory,
except that \n{packing(e)} stands for `packing by element' and
\n{packing(v)} for `packing a vector data type'.

The graphs show for each installation:
\begin{itemize}
\item The measured time of one ping-pong, as function of message size
  in bytes;
\item the effective bandwidth; and
\item the slowdown with respect to the contiguous send. The value of
  this plot is mostly for smaller messages, where differences between
  schemes are not apparent in the first two plots. Note that
  occasional
  blips in the reference curve are reflected in the slowdown numbers.
\end{itemize}
We start by discussing general tendencies,
concluding with the relatively minor differences between the various installations.

\subsection{Derived datatypes}

We see that in most cases the time for sending a derived datatype
tracks the time for sending a manually copied buffer very well. This
can best be explained by assuming  that sending a derived datatype is done by copying the
data internally to contiguous storage, which is then sent.

However, we see a drop in performance for messages beyond a few tens
of megabytes. We assume that for such relatively large messages the
internal buffer bookkeeping of MPI becomes complicated, and incurs
extra overhead.

Since many MPI implementations are closed source, and the behaviour
depends on the precise scheme, we can not further analyze this.
(In fact, even for
the open MPICH implementation we found the buffer management code too
inscrutable for analysis.)

\subsection{Buffered sends}

MPI buffered sends (that is, passing a userspace buffer with
\n{MPI_Buffer_attach} for MPI to use internally, and calling
\n{MPI_Bsend}) are supposed to alleviate limitations of MPI's internal buffering.
Strangely, having a fully allocated userspace buffer does not help the
slowdown for large messages. In fact, in most MPI implementations it
performs worse, even for intermediate message sizes.

\subsection{Packed buffers}

The elementwise packing scheme performs predictably very badly.
However, packing a derived type gives essentially the same
performance as manual copying. This proves two points:
\begin{itemize}
\item The \n{MPI_Pack} command is as efficient as a user-coded copying
  loop.
\item  There are
  no intrinsic inefficiencies in the derived datatype code.
\end{itemize}
The fact that this scheme performs better than using a derived type
directly also indicates that there are inefficiencies in the internal
buffer management of MPI that are obviated by having the buffer
totally in user space.

\subsection{One-sided communication}

The performance of one-sided communication does not track the previous
schemes. Its behaviour seems dependent on the message size.
\begin{enumerate}
\item For small messages, one-sided transfer is slow, probably because
  of the more complicated synchronization mechanism of
  \n{MPI_Win_fence}, which imposes a large overhead.
\item For intermediate size messages, one-sided transfer is
  competitive with other schemes, except for the mvapich
  implementation where it is several factors slower.
\item For large messages the range of behaviours is even
  larger. However, there are few cases where one-sided communication
  is truly competitive.
\end{enumerate}

\subsection{Eager limit}

Most MPI implementations have a switch-over between the `eager
protocol', where messages are send without handshake, and a
`rendezvous' protocol that does involve acknowledgement.
From its nature, the eager protocol relies on sufficient buffer space,
so there is an `eager limit'. One typically observes that messages
just over the eager limit can perform worse (at least per byte) than
once just under.

The effects of the eager limit are visible in most plots, with the
following remarks:
\begin{itemize}
\item As is to be expected, we mostly see a performance drop at the
  eager limit for all send schemes. For one-sided puts it is less
  pronounced, and for the packing scheme it largely drowns in the
  overhead.
\item On Cray-mpich we see the performance drop for the reference
  speed, and at double the data sizes for the packing scheme. For the
  other schemes not much of a drop is visible. The reason for this is
  unclear.
\end{itemize}

We have tested setting the eager limit over the maximum message size,
but this did not appreciably change the results for large messages.

\subsection{Cache flushing}


In tests not reported here we dispensed with flushing the cache in
between sends. This had a clear positive effect on intermediate size messages.

\subsection{Further tests}

This paper is of course not a full exploration of all possible
scenarios in which derived datatypes can be used. In fact, it only
examines the very simplest case of a derived type.
\begin{enumerate}
\item Types with less regular spacing may give worse performance due
  to decreased use of prefetch streams in reading data.
\item Types with larger block sizes may perform better due to higher
  cache line utilization in the read.
\end{enumerate}
A more interesting objection to these tests is that we used exactly
one communicating process per node. This is a realistic model for
certain hybrid cases, but not for `pure MPI' codes. However, a limited
test, not reported in graph form here, shows that no performance
degradation results from having all processes on a node
communicate. (In fact, sometimes a slightly higher bandwidth results.)

The interested reader can find the code in an open repository and
adapt it to their own purposes~\cite{packingrepo}.

\subsection{Differences between installations}

We give the results on Skylake nodes with OmniPath fabric and Intel
MPI (henceforth: \n{impi}) as representative in
figure~\ref{fig:skx-i}. Switching to \n{mvapich2} gives largely the
same results; figure~\ref{fig:skx-v}.

The native Cray MPI also has similar performance, with the 
exception that one-sided performance for large sizes is on par with the derived types,
unlike on Stampede2 where for all sizes it shows a relative degradataion; figure~\ref{fig:ls5}.

Finally, Stampede2-knl shows the same peak network performance, but
the performance of our non-contiguous tests is hampered by 
the core performance in constructing the send buffer; figure~\ref{fig:knl}.

\section{Conclusion}
\label{sec:conclusion}

Schemes for sending non-contiguous data between two processes are
considerably slower than sending contiguous data. The slowdown of at
least a factor of three can mostly be explained by multiple memory
reads and a lack of overlap of memory and network traffic.
Attaining such overlap for non-contiguous data depends on advanced
functionality of the network interface~\cite{LI:MpiDataUMR}.

For any but large (over $10^8$ bytes) messages the various schemes
perform fairly similarly, so there should be no reason not to use
derived datatypes, these being the most user-friendly. Depending on
the architecture, one-sided transfers may behave worse, and buffered
sends are at a disadvantage, both even at intermediate sizes
($10^4$--$10^8$ bytes).

The scheme that consistently performs best applies \n{MPI_Pack} to a
derived datatype. This scheme in all cases gives the same performance
as a manual gather copy of the data.

\bibliographystyle{plain}
\bibliography{vle}

\end{document}